\documentclass[global,twocolumn,a4paper]{svjour}
\usepackage{graphicx}
\usepackage{psfrag}
\journalname{}
\begin{document}
\title{An all-optical ion-loading technique for scalable microtrap architectures}
\author{R.J. Hendricks, D.M. Grant, P.F. Herskind, A. Dantan, M. Drewsen}                  
\institute{QUANTOP --- Danish National Research Foundation Center for Quantum Optics, Department of Physics and Astronomy, University of Aarhus, DK-8000 \AA rhus C., Denmark}
\date{}
%
%
%
\maketitle
%
%
\begin{abstract}
An experimental demonstration of a novel all-optical technique for loading ion traps, that has particular application to microtrap architectures, is presented.  The technique is based on photo-ionisation of an atomic beam created by pulsed laser ablation of a calcium target, and provides improved temporal control compared to traditional trap loading methods.  Ion loading rates as high as 125 ions per second have so far been observed.  Also described are observations of trap loading where Rydberg state atoms are photo-ionised by the ion Doppler cooling laser.
\end{abstract}
%
%
\section{Introduction}
\label{sec:introduction}
In recent years, a number of experiments have demonstrated ion traps to be a viable technology for quantum information processing (\textrm{QIP}) and direct quantum simulation~\cite{Leibfried03,Haffner05,NIST05:6ionGHZ}.  There are, however, a number of obstacles that must still be overcome if ion trap \textrm{QIP} is to solve problems that are intractable using classical computing methods.  

One of the most important of these obstacles is the scaling up of current experiments to enable processing of much larger numbers of ion qubits.  As a result, there has been much recent interest in the development of microscale arrays of ion traps~\cite{Kielpinski02,Wineland05,Stick06,Brownnutt06}.

Besides the challenges involved in fabrication of microtrap arrays, there are a number of problems associated with reducing the size of ion traps.  One of the most significant of these is the observation that the rate at which ions are heated has been measured to increase dramatically as the trap size is reduced.  Experiments have measured the heating rate to vary with trap scale, $r$, as strongly as $r^{-4}$~\cite{Michigan06:needletrapheating}.  This increase is attributed to patch potentials due to the presence of contaminants on the trap electrodes.  Often this contaminant material is deposited by the atom sources used to load ion traps~\cite{Devoe02}.  An interesting observation is that the effect of these patch potentials can be significantly suppressed by cooling of the trap electrodes to cryogenic temperatures.

At present, most ion traps are loaded by electron bombardment or photo-ionisation of a neutral atomic beam~\cite{Kjaergaard00}.  Atoms are typically emitted from a resistively heated oven, and a series of skimmers is used to produce from this source a nominally collimated atom beam which is directed towards the centre of the trapping region.  This beam is continuous and cannot be rapidly switched on or off, so many more atoms are sent through the trap than are required.  Over time this can lead to the build up of material on trap electrodes and nearby surfaces.  A cleaner ion-loading technique would result in less build up of contaminant material on electrodes and therefore reduce the extent of heating as trap size is reduced.  An atom source that can be more rapidly turned on and off to provide a better controlled atom flux than a traditional thermal source is therefore highly desirable.

Here we present a general all-optical technique for loading ion traps that has a number of advantages over thermal atom sources.  In our experiment a pulsed laser source is used to ablate material from a calcium metal surface.  The atoms produced are directed through the trap, where they are photo-ionised.  Through rapid swit\-ching on and off of the ablation laser, atoms can be produced in brief pulses only when required~\cite{Nogar85}.  This reduces the overall amount of material passing through the trap and so reduces electrode contamination.  The pulsed nature of the technique also makes it suitable for the controlled production of single ions, as required for scalable \textrm{QIP}.  The general technique can be scaled down in size in order to better integrate with microtrap structures, can be used with many different atomic species and is in principle compatible with operation at cryogenic temperatures.
%
%
\section{Laser ablation}
\label{sec:laserablation}
Laser ablation is a diverse and complex field and the reader is directed to references~\cite{Ashfold04,Phipps} for an overview.  In brief, there are two distinct regimes in which nanosecond pulse laser ablation occurs, distinguished largely by the laser fluence (energy deposited on a surface per unit area by each pulse).  Here we are principally concerned with the low fluence regime, in which thermal processes are dominant.  If the laser pulse duration is much shorter than the timescale for thermal conduction processes then the region of the target that is irradiated by the laser may be locally heated to high temperatures~\cite{Balazs91}.  This can result in melting, sublimation or desorption of material in the affected region.  Laser ablation in this thermal regime has previously been used to reduce the amount of source material required for resonance ionisation mass spectrometry~\cite{Nogar85}.  The motivation is therefore similar to that described in the present work.

As the laser fluence is increased, multi-photon excitation will eventually lead to ionisation of material at or near the target surface, resulting in the formation of a plasma phase~\cite{Kelly98}.  This can dramatically increase the rate at which energy is absorbed from the laser beam.  In this high fluence regime there are a variety of mechanisms by which material can be ejected from the target surface, some of which occur on a macroscopic scale~\cite{Ashfold04}.  For the purposes of loading ion traps the plasma regime may sometimes be undesirable because a significant number of excited state atoms can be produced.
%
%
\section{Experimental techniques}
\label{sec:experimental}
In our experimental setup we trap $^{40}$Ca$^+$ ions.  These ions are laser cooled on the rapid S$_{1/2} \longrightarrow$ P$_{1/2}$ dipole transition at 397nm.  A repumper at 866nm is required in order to prevent build up of population in the metastable D$_{3/2}$ state~\cite{Mortensen04}.  Ions are detected by imaging the spontaneously re-emitted 397nm light onto an image intensifier, the output of which is monitored with a \textrm{CCD} camera.  The fluorescence rate on the 397nm transition is such that, given the magnification in the optical system of approximately 10, it is possible to detect and resolve individual ions.  

The ion trap used in these test experiments is a linear radiofrequency trap that has been described in detail elsewhere~\cite{Mortensen05}.  It has a trap size parameter, $r$, of 2.35mm and is operated at a frequency of 4.0MHz.  In order to reduce the rate of collisions between the ions and background gas particles, the trap is housed in an ultra-high vacuum chamber at a base pressure of approximately $4 \times 10^{-10}$mbar, as measured on an ion gauge within the chamber.

With appropriate trapping voltages it is possible to form linear strings of ions or Coulomb crystals containing rather large numbers of ions~\cite{Drewsen98,Mortensen06}.  The number of ions in such crystals can be determined by measuring the overall crystal volume and using the trapping parameters to calculate the ion density~\cite{Mortensen06,Madsen00}. 

\begin{figure}
\begin{center}
\includegraphics[width=0.45\textwidth]{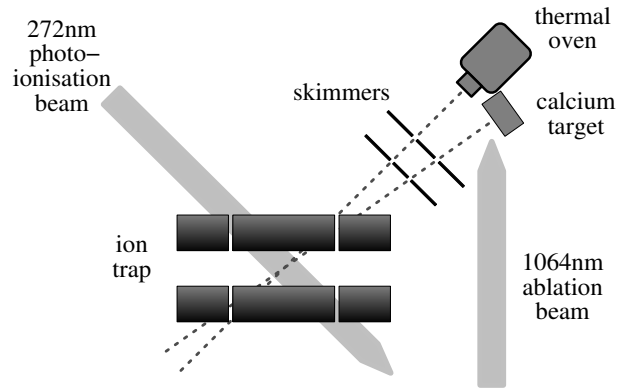}
\caption{Ion-loading setup within the trap chamber.  Atoms are produced either from a thermal source or from a calcium target irradiated with light from a pulsed Nd:YAG laser.  A series of skimmers prevents any of this material from contaminating the trap electrodes.  Atoms which reach the centre of the trap can be photo-ionised by 272nm light directed almost perpendicular to the atomic beams.
\label{fig:trapsetup}}
\end{center}
\end{figure}

Usually the trap is loaded using atoms from a thermal source (described in~\cite{Kjaergaard00,Mortensen05}).  A series of skimmers is used to trim the distribution of emitted atoms into a narrow beam which is approximately 1.0mm $\times$ 1.5mm at the centre of the trap (see figure~\ref{fig:trapsetup}).  The thermal source is about 13cm from the trap, so a rather small fraction of the atoms produced passes through the skimmers.
Atoms passing through the centre of the trap can be photo-ionised in a two-photon process by light at 272nm.  The first step in this process is a resonant transition and hence is isotope selective.  The relevant energy levels are shown on figure~\ref{fig:energylevels} and full details can be found in~\cite{Kjaergaard00,Mortensen04}.  The 272nm light is generated by frequency-quadrupling light from a fibre laser at 1088nm~\cite{Herskind07}.  Typically the 272nm power used is about 15mW, and the beam waist at the centre of the trap is approximately 160$\mu$m.  In order to reduce Doppler shifting of the ionisation laser frequency due to the velocity of the atoms in the beam, the propogation direction of the laser is chosen to be perpendicular to that of the atom beam.

In the present work, a pure calcium metal target is placed close to the thermal source with an identical set of skimmers leading to the centre of the trap (see figure~\ref{fig:trapsetup}).  The surface of the calcium target is aligned such that it is perpendicular to the path towards the trap centre.  Due to geometric constraints, this path is aligned at an angle of about 12 degrees relative to the ionisation laser.  

The ablation laser used in these experiments is a 1064nm pulsed Nd:YAG laser (CrystaLaser QIR-1064-500).  The maximum pulse energy is about 80$\mu$J and remains constant for pulse repetition rates up to about 3kHz.  At repetition rates higher than this the maximum energy begins to decline, and by about 15kHz it falls as the inverse of the repetition rate.  Pulse durations are also dependent on the repetition rate, but are in the region of 30-50ns for the rates used here.  The maximum repetition rate is 200kHz.    The laser is focussed onto the calcium target with a waist of about 75$\pm$15$\mu$m.  In our current setup we are restricted to using an angle between the laser propogation direction and the normal to the calcium surface of approximately 30 degrees.  This means that the region illuminated by the ablation laser will be slightly larger in one direction than the measured beam waist.

\begin{figure}
\begin{center}
\psfrag{30}{\large30$^{\circ}$}
\includegraphics[width=0.45\textwidth]{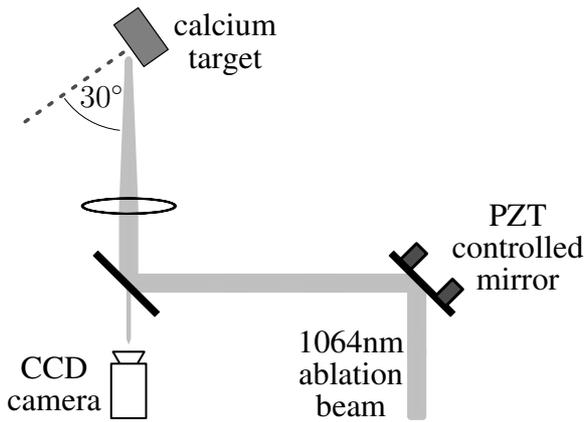}
\caption{Setup for alignment and control of the ablation laser beam.  The \textrm{CCD} camera images the target surface, and is used for initial laser alignment.  By filtering out light at 1064nm it can also be used to monitor the extent of plasma formation at the target.  The \textrm{PZT} controlled mirror is used to dither the beam position in the vertical and horizontal directions.
\label{fig:alignment}}
\end{center}
\end{figure}

Alignment of the ablation laser is performed by imaging the target with a \textrm{CCD} camera directed parallel to the beam path~(see figure~\ref{fig:alignment}).  With a low pulse energy but large repetition rate, scattered light from the calcium target is clearly visible and allows positioning of the beam.  By placing a filter in front of the \textrm{CCD} camera, this light at 1064nm is eliminated.  The pulse energy can then be gradually increased until light from the calcium target is once again seen on the camera.  This light at a wavelength other than 1064nm is a clear indication of the generation of plasma on the calcium surface.  The focussing of the ablation laser beam can then be optimised by finding the point at which plasma is observed with the lowest pulse energy.  Once the position of focus has been optimised, we find that plasma can be clearly seen with ablation laser fluences above about 600mJ/cm$^2$.

To ensure that the ablation beam covers the region of the calcium target that is aligned with the skimmers and the centre of the trap, a \textrm{PZT} controlled mirror is used to dither the beam position over a region of about 1mm$^2$.  We have found that this also helps to keep the ablation rate stable.

In addition to the main ion trap setup, we have also constructed a simple test chamber for characterising the ablation process.  In this chamber we can use larger targets and can rapidly replace them after use.  This means that we can ablate much larger quantities of material and can use targets with a much greater visible surface area, so that we can ablate many holes using different laser parameters.  The test chamber is pumped down to a pressure of about $1 \times 10^{-5}$mbar.  The alignment procedure and beam dithering technique are the same as those used in the ion trap setup described earlier.  In this setup the ablation laser waist is approximately 55$\pm$15$\mu$m and the propogation direction is almost normal to the calcium surface.  Due to uncertainties in beam waist measurements and the rather different angles of incidence, care should be taken when comparing fluence measurements in the two setups.

To determine the quantity of material that has been ablated from the target in this setup, we remove the target and use optical microscopy to measure the dimensions of the ablation craters.  For this technique to be effective we must necessarily remove material to a depth of at least a few micrometers.
%
%
\section{Results}
\label{sec:results}
\begin{figure}
\begin{center}
\begin{tabular}{c}
\large\textbf{(a)}
\\
\includegraphics[width=0.45\textwidth]{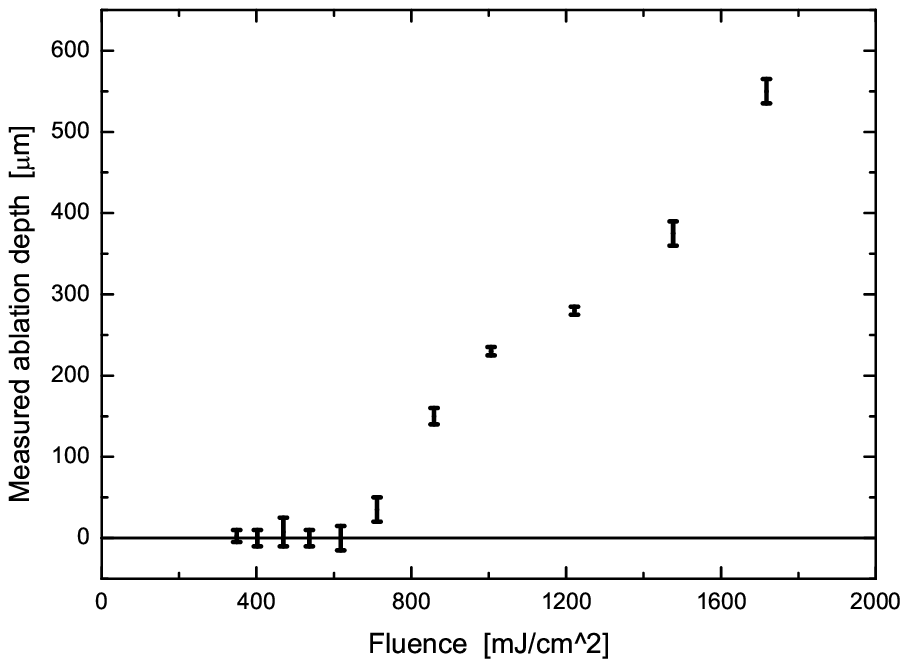}
\\
\\
\large\textbf{(b)}
\\
\includegraphics[width=0.45\textwidth]{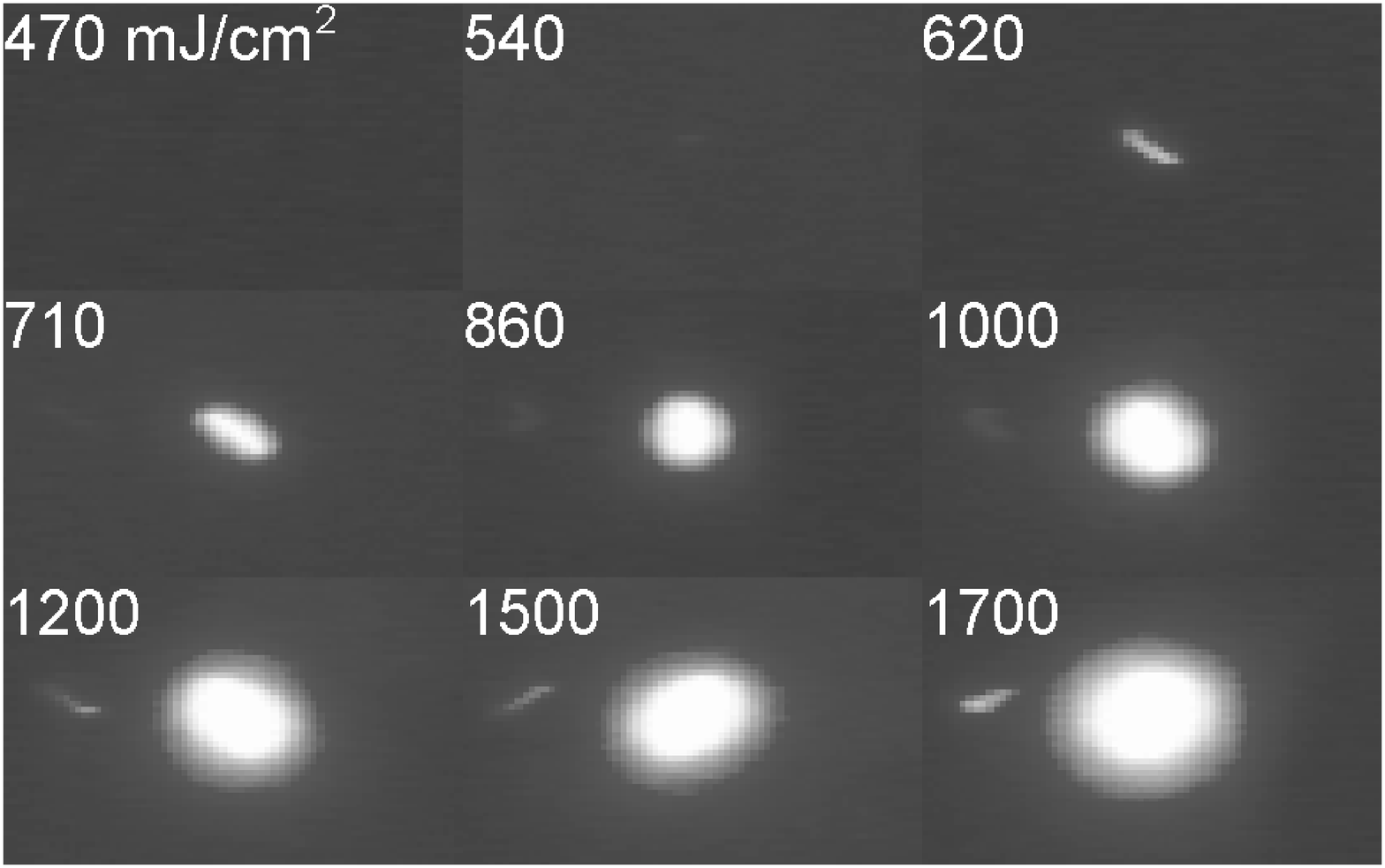}
\end{tabular}
\caption{\textbf{\textit{(a)}}~Depth of holes ablated in a calcium target after 4.6 million pulses, plotted as a function of laser fluence.  The bars indicate the measured maximum and minimum depth of the ablated surface, relative to the original surface.  The observed threshold behaviour at about 600mJ/cm$^2$ appears to correspond to the onset of plasma formation.\newline\textbf{\textit{(b)}}~\textrm{CCD} images of the target showing the extent of the plasma for different ablation laser fluences (marked on the images in mJ/cm$^2$).  Each image corresponds directly to a plotted point in (a).
\label{fig:ablationvsfluence}}
\end{center}
\end{figure}

Using a calcium target in the test chamber, the ablation depth as a function of pulse energy has been studied using in each case 4.6 million pulses fired at a repetition rate of 2.4kHz.  The results are plotted in figure~\ref{fig:ablationvsfluence}a.  The bars in this plot indicate the maximum and minimum depth of the ablation hole relative to the original calcium surface.  With fluences less than or equal to about 600mJ/cm$^2$ there are regions that are actually raised relative to the initial surface and it is impossible to precisely determine how much material has been removed.  What is seen using optical microscopy is a series of raised and pitted regions that seems to be due to localised melting and redistribution of the metal surface.  This would be consistent with a very low ablation rate resulting from a thermally driven process.  It would appear from figure~\ref{fig:ablationvsfluence}a that there is a threshold fluence at which the ablation rate begins to increase much more steeply, and almost linearly with  the laser fluence.  This would seem to correspond to the onset of plasma generation and hence the presence of a second, much more rapid, ablation mechanism.  This is supported by the direct observation of plasma formation at around this fluence, using a \textrm{CCD} camera with strong filtering of the scattered light at 1064nm.  Typical images taken with this camera at different ablation laser fluences are indicated in figure~\ref{fig:ablationvsfluence}b.  The faint second image to the left of each plasma plume is due to double reflection in the beamsplitter placed in front of the \textrm{CCD} camera.

It is important that atoms are emitted in their electronic ground states if the photo-ionisation process is to remain isotope-selective.  In practice this means that ablation must be carried out in the thermal regime, where plasma effects are not observed.  Note also that given the extremely small atom flux required for loading an ion trap, it is anticipated that the lifetime of a calcium target can be very long.

\begin{figure}
\begin{center}
\includegraphics[width=0.45\textwidth]{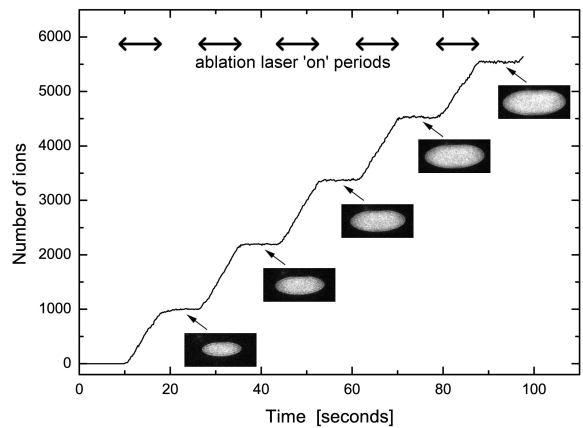}
\caption{Number of ions present in the trap as a function of time, when the ablation laser is periodically gated on and off.  The ion number is seen to increase only during the `on' phase of the gating cycle, indicated by the double-headed arrows.  Ion number is determined from analysis of the dimensions of the ion crystal in the trap at each point.  The inset images are examples that show the state of the crystal at the indicated times.  The ablation laser fluence used is 240mJ/cm$^2$and the repetition rate is 25kHz.
\label{fig:gatedloading}}
\end{center}
\end{figure}

In the ion trap setup, ions can be loaded into the trap by irradiating the calcium target with the ablation beam whilst the photo-ionising laser is operating continuously.  Figure~\ref{fig:gatedloading} shows the calculated number of ions present in the trap as a function of time, whilst the ablation laser is periodically gated with regular `on' and `off' periods of nine seconds. The number of ions is seen to increase during the `on' periods.  This increase stops abruptly whenever the laser is switched off, indicating that ion loading is indeed due to ionisation of calcium atoms emitted by the ablation process.

The loading rate is found to be dependent on the ablation laser fluence.  Rates as high as 125 ions per second have so far been observed, using a fluence of 240mJ/cm$^2$ at a repetition rate of 25kHz.  There are many other parameters that can affect the loading rate, such as the ablation beam positioning and the photo-ionisation laser wavelength and intensity.  It is expected that with further optimisation even greater loading rates could be achieved, whilst still keeping the ablation laser fluence well below that required to generate plasma.

\begin{figure}
\begin{center}
\includegraphics[width=0.45\textwidth]{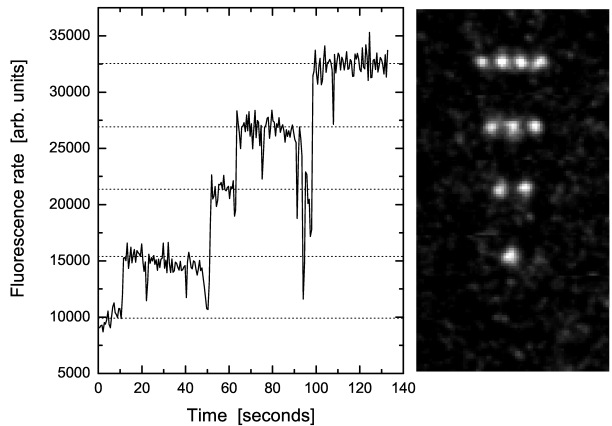}
\caption{Fluorescence detected from the centre of the trap as a function of time during continuous ablation with a laser fluence of 120mJ/cm$^2$ and repetition rate of 50kHz.  With these parameters the loading rate is extremely small and single ions can easily be loaded.  The intensified \textrm{CCD} images show the ion string shortly after each ion is loaded.
\label{fig:singleionloading}}
\end{center}
\end{figure}

In order to demonstrate the ability of the ablation loading technique to load individual ions in a controlled manner, experiments have been carried out with laser fluences as low as 120mJ/cm$^2$.  In this regime ions are loaded rather slowly, and it is a straightforward task to shutter the ablation laser once a desired number of ions is reached.  Figure~\ref{fig:singleionloading} shows a number of frames taken from a series of \textrm{CCD} images recorded during continuous ion loading at this laser fluence.  The number of ions present in the string is clearly apparent in each frame.  Also shown in figure~\ref{fig:singleionloading} is the overall fluorescence as determined by integrating the signal in a fixed area of each of the \textrm{CCD} images.  The time at which each of the ions is loaded is clearly marked by a discrete step in the fluorescence level.  Brief drops in fluorescence are sometimes observed shortly before a new ion is observed.  These drops are due to heating of the ions already in the trap by the new, much hotter ion.  Drops in signal are sometimes seen at other times, when an ion is excited into a non-fluorescing state through collisions with atoms passing through the trapping region.  

Although the rate at which ions are loaded in this example is intentionally rather small, it would be relatively straightforward to implement a loading scheme whereby the fluorescence rate is continually monitored and the ablation laser is automatically shuttered when an ion is loaded.  With such a system, individual ions could be controllably loaded at higher fluences and hence much more rapidly.  

\begin{figure}
\begin{center}
\includegraphics[width=0.45\textwidth]{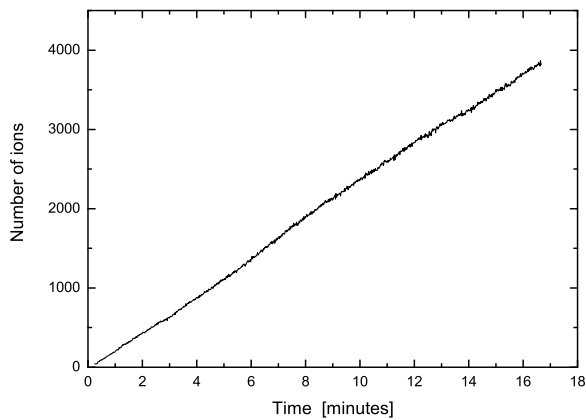}
\caption{Number of ions present in the trap as a function of time during continuous ablation with a fluence of 240mJ/cm$^2$ and repetition rate of 25kHz.  The photo-ionisation parameters are set such that the loading rate observed is rather small, but it is seen to be almost constant for a period of more than fifteen minutes.
\label{fig:linearloading}}
\end{center}
\end{figure}

In order to achieve well controlled loading of the trap, it is important that the loading rate be stable over time.  That this is the case is demonstrated in figure~\ref{fig:linearloading}, which shows the number of ions in an ion crystal during more than fifteen minutes of continual loading.  The very slight oscillation in the loading rate can probably be attributed to drift in the wavelength of the photo-ionising laser, since similar effects are seen during loading with a thermal source.

An additional benefit of the laser ablation-based ion-loading technique is that the pressure rise during loading can be kept relatively small.  Thermal atom sources must necessarily be maintained at some high temperature during operation.  This raised temperature often leads to a significant increase in the background pressure within the vacuum chamber.  Even after loading is completed it can take several minutes for the pressure to return to its initial level.  During laser ablation, the surface area of the target that is strongly heated is governed by the size of the ablation beam.  Since this can be very small, it is possible to avoid large pressure rises during loading and obtain much more rapid recovery times.

Upon first ablating the surface of the calcium target in our ion trap setup, the pressure was observed to rise as high as $10^{-8}$mbar.  This initial increase is ascribed to evaporation of contaminant material on the surface layer of the target.  As ablation continued, much of this contaminant layer was removed and the pressure in the chamber improved.

\begin{figure}
\begin{center}
\includegraphics[width=0.45\textwidth]{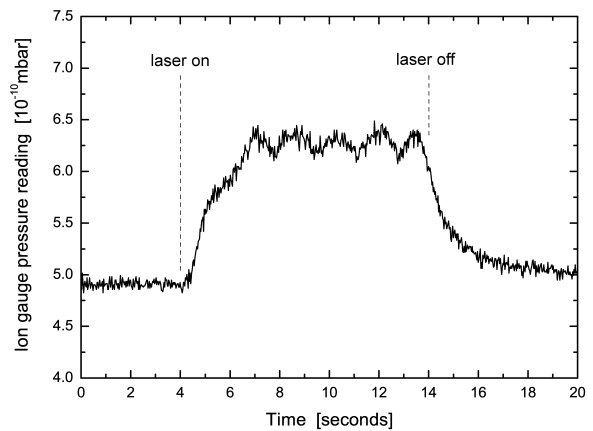}
\caption{Response of ion gauge pressure measurement to a ten second period of ablation loading with a laser fluence of 270mJ/cm$^2$ and repetition rate of 23kHz.  An increase in the equilibrium pressure of about $1.5 \times 10^{-10}$mbar is observed.  The pressure returns to its initial value within a few seconds of switching off the ablation laser.
\label{fig:pressureresponse}}
\end{center}
\end{figure}

The typical pressure response during ablation of the calcium target with a relatively large laser fluence of 270mJ/cm$^2$ and a repetition rate of 23kHz is shown in figure~\ref{fig:pressureresponse}.  The ablation continues for a period of ten seconds, during which the pressure rises to a new equilibrium level that is approximately $1.5 \times 10^{-10}$mbar greater than the base pressure.  The pressure returns to its initial value within a few seconds of switching off the ablation laser.  It is possible that the pressure rise will be further reduced over time as more contaminants are removed from the target surface.
%
%
\section{Photoionisation of Rydberg atoms}
\label{sec:directloading}
Besides the trap loading technique described in the previous sections, we have also observed that at relatively high ablation laser fluences ions can be loaded into the trap without the use of the photo-ionisation laser at 272nm.  The loading rates for this process can be as high as 25 ions per seconds.  Although it is not fully understood how the various experimental parameters affect the loading rate we do find that at relatively low fluences it is possible to eliminate this loading process altogether.  Indeed, during collection of all the data presented in the previous section no loading was observed without the 272nm laser.

This loading of the trap without the 272nm laser is surprising, especially since it occurs even if the 866nm repumper laser is blocked.  This eliminates the laser cooling force and makes it extremely unlikely that an ion generated outside the trap should be captured.  It follows that the ions must be generated near the centre of the trap as a result of photo-ionisation by the Doppler cooling or repumper lasers.  By blocking each of these beams in turn, it has been determined that it is only the 397nm cooling beam that is responsible for this photo-ionisation.

\begin{figure}
\begin{center}
\includegraphics[width=0.45\textwidth]{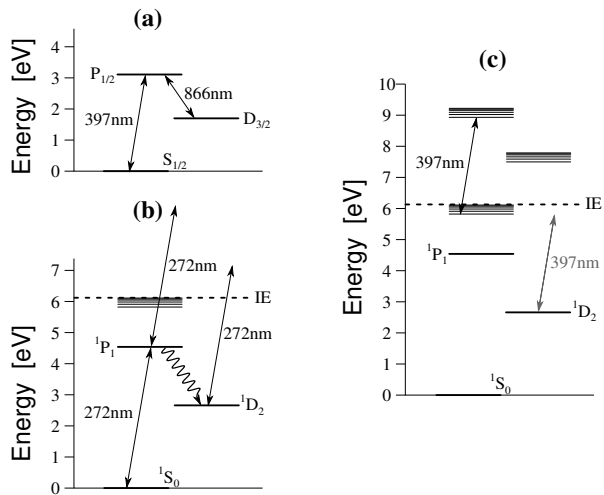}
\caption{\textbf{\textit{(a)}}~Partial energy level diagram of Ca$^+$.\newline  \textbf{\textit{(b)}}~Partial energy level diagram of atomic calcium.\newline  \textbf{\textit{(c)}}~By superimposing the ionic energy levels onto each of the atomic Rydberg levels, we obtain approximate values for some of the doubly-excited auto-ionising states of atomic calcium.
\label{fig:energylevels}}
\end{center}
\end{figure}

Since it takes some time for the ablated atoms to reach the centre of the trap, they must necessarily be produced in some long-lived excited electronic state.  The most obvious candidate would perhaps be the metastable 4s3d~$^1$D$_2$ state, but the photon energy at 397nm is insufficient to ionise from this level (see figure~\ref{fig:energylevels}c).  Alternatively it is possible that the atoms are produced in bound Rydberg states, from which resonant transitions to some auto-ionising atomic states exist at 397nm but not at 866nm.

The outer electron in a high-lying Rydberg state atom can be viewed as a spectator that does not significantly interact with more tightly bound electrons.  The transitions that are available to an atom in such a Rydberg state are therefore very similar to those available to the primary ion.  The energy levels of the doubly-excited states can be approximated by superimposing the ionic energy level system onto each of the singly-excited Rydberg levels (see figure~\ref{fig:energylevels}).  It follows that regardless of the exact initial Rydberg state of the calcium atoms there will always be a transition to an auto-ionising state at close to the 397nm ion cooling wavelength.  The validity of this approximation is confirmed by detailed calculations and measurements of such an auto-ionising series in calcium~\cite{Bolovinos96}.

\begin{figure}
\begin{center}
\includegraphics[width=0.45\textwidth]{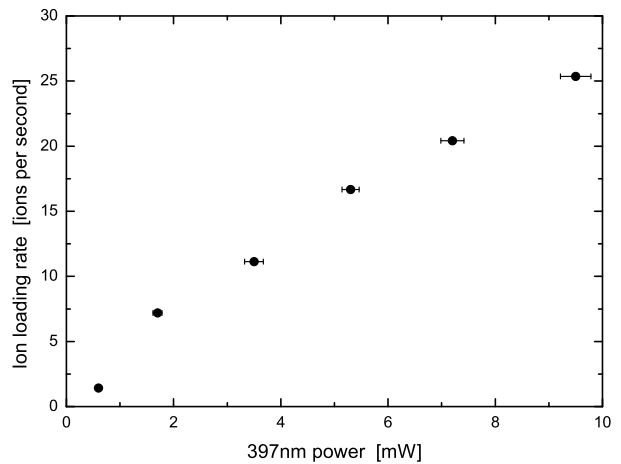}
\caption{Ion loading rate as a function of 397nm laser power.  An initially linear relationship is observed, with possible saturation effects beginning to occur at higher powers.  The ablation laser fluence used is 300mJ/cm$^2$, with a repetition rate of 20kHz.
\label{fig:397power}}
\end{center}
\end{figure}

We have studied the rate at which ions are loaded by this process as a function of 397nm laser power.  The results are presented in figure~\ref{fig:397power}.  An initially linear relationship is observed that shows some signs of saturation at higher powers.

Since the auto-ionising resonances are rather broad, this trap loading technique is not expected to be isotope selective and is therefore of limited use compared to resonant photo-ionisation of ground state calcium atoms.  We note, however, that the same loading effect would be expected for any element that has a rapid Doppler cooling transition available to it in the ionic state.  The technique may therefore be particularly useful for elements that only possess one isotope, such as beryllium, or where isotope selectivity is not required.  The fraction of atoms produced in the Rydberg states, and hence the overall efficiency of this loading process, has yet to be determined.
%
%
\section{Outlook}
\label{sec:outlook}
The general technique for ion trap loading presented here provides improved temporal control over the loading process, when compared to traditional methods.  This may make it possible to reduce the overall amount of material passing through the trap, and hence to alleviate the problem of electrode contamination

The technique can in principle be scaled down so that the target and laser delivery systems are integrated into microtrap structures.  They would then form the basis of a `loading zone', providing a stream of ions for a microtrap array.  High-fluence laser ablation has previously been used for time-resolved studies of trapped, laser-cooled atoms~\cite{Raab87}.  The low-fluence ablation loading technique presented here may also be well suited to loading of the miniature atomic chip traps currently being developed~\cite{Reichel02}.

Because the mean power dissipated in the ablation process is small, the loading technique presented here is compatible with traps operated at cryogenic temperatures.  This feature may also be of interest to other parts of the ion and atom trapping community.  Some species are not often trapped because very high temperatures are necessary to generate the required atomic beams.  Chromium cells, for example, are often heated  to almost 2000K to achieve a sufficient vapour pressure for trapping~\cite{Bradley00,Griesmaier05}.  Maintaining a thermal source at such high temperatures makes it technically demanding to obtain low pressures in the vacuum chamber.  By using a pulsed laser heating technique this problem might be alleviated.

Finally we note that the photo-ionisation of Rydberg state atoms using ion Doppler cooling transitions may be of benefit in ion trap experiments where isotope selectivity is not required, since it provides a simple route to some of the benefits of more complicated photo-ionisation schemes.
%
%
\section*{Acknowledgements}
\label{sec:acknowledgements}
The authors would like to thank Jacques Chevalier and Folmer Lyckegaard for technical assistance and numerous helpful discussions.  This work is financially supported by the Carlsberg Foundation and by the EU under contract IST-517675-MICROTRAP.
%
%
\bibliographystyle{unsrt}
\bibliography{ablation_loading_paper}

\begin{thebibliography}{10}

\bibitem{Leibfried03}
D.~Leibfried, R.~Blatt, C.~Monroe, and D.~Wineland.
\newblock Quantum dynamics of single trapped ions.
\newblock {\em Reviews of Modern Physics}, 75:281--324, 2003.

\bibitem{Haffner05}
H.~H\"affner et~al.
\newblock Scalable multiparticle entanglement of trapped ions.
\newblock {\em Nature}, 438:643--646, 2005.

\bibitem{NIST05:6ionGHZ}
D.~Leibfried et~al.
\newblock Creation of a six-atom `{S}chr\"odinger cat' state.
\newblock {\em Nature}, 438:639--642, 2005.

\bibitem{Kielpinski02}
D.~Kielpinski, C.~Monroe, and C.~Wineland.
\newblock Architecture for a large-scale ion-trap quantum computer.
\newblock {\em Nature}, 417:709--711, 2002.

\bibitem{Wineland05}
D.J. Wineland et~al.
\newblock Quantum control, quantum information processing, and quantum-limited
  metrology with trapped ions.
\newblock In {\em Proc. 17th Int. Conf. Laser Spect.}, pages 393--402, 2005.

\bibitem{Stick06}
D.~Stick, W.K. Hensinger, S.~Olmschenk, M.J. Madsen, K.~Schwab, and C.~Monroe.
\newblock Ion trap in a semiconductor chip.
\newblock {\em Nature Physics}, 2:36--39, 2006.

\bibitem{Brownnutt06}
M.~Brownnutt, G.~Wilpers, R.C. Thompson, and A.G. Sinclair.
\newblock Monolithic microfabricated ion trap chip design for scaleable quantum
  processors.
\newblock {\em New Journal of Physics}, 8:232, 2006.

\bibitem{Michigan06:needletrapheating}
L.~Deslauriers, S.~Olmschenk, D.~Stick, W.K. Hensinger, J.~Sterk, and
  C.~Monroe.
\newblock Scaling and suppression of anomalous heating in ion traps.
\newblock {\em Physical Review Letters}, 97:103007, 2006.

\bibitem{Devoe02}
R.G. DeVoe and C.~Kurtsiefer.
\newblock Experimental study of anomalous heating and trap instabilities in a
  microscopic ${}^{137}${B}a ion trap.
\newblock {\em Physical Review A}, 65:063407, 2002.

\bibitem{Kjaergaard00}
N.~Kj{\ae}rgaard, L.~Hornek{\ae}r, A.M. Thommesen, Z.~Videsen, and M.~Drewsen.
\newblock Isotope selective loading of an ion trap using resonance-enhanced
  two-photon ionization.
\newblock {\em Applied Physics B}, 71:207--210, 2000.

\bibitem{Nogar85}
N.S. Nogar, R.C. Estler, and A.M. Miller.
\newblock Pulsed laser desorption for resonance ionization mass spectrometry.
\newblock {\em Analytical Chemistry}, 57:2441--2444, 1985.

\bibitem{Ashfold04}
M.N.R. Ashfold, F.~Claeyssens, G.M. Fuge, and S.J. Henley.
\newblock Pulsed laser ablation and deposition of thin films.
\newblock {\em Chemical Society Reviews}, 33:23--31, 2004.

\bibitem{Phipps}
C.R. Phipps, editor.
\newblock {\em Laser Ablation and its Applications}.
\newblock Springer, 2007.

\bibitem{Balazs91}
L.~Balazs, R.~Gijbels, and A.~Vertes.
\newblock Expansion of laser-generated plumes near the plasma ignition
  threshold.
\newblock {\em Analytical Chemistry}, 63:314--320, 1991.

\bibitem{Kelly98}
R.~Kelly et~al.
\newblock Plume formation and characterization in laser-surface interactions.
\newblock In J.~Miller and R.F. Haglund, editors, {\em Laser Ablation and
  Desorption}, pages 225--289. Academic Press, 1998.

\bibitem{Mortensen04}
A.~Mortensen, J.J.T. Lindballe, I.S. Jensen, P.~Staanum, D.~Voigt, and
  M.~Drewsen.
\newblock Isotope shifts of the {$4s^2 \; {}^1S_0 \rightarrow 4s5p \; {}^1P_1$}
  transition and hyperfine splitting of the {$4s5p \; {}^1P_1$} state in
  calcium.
\newblock {\em Physical Review A}, 69:042502, 2004.

\bibitem{Mortensen05}
A.~Mortensen.
\newblock {\em Aspects of Ion Coulomb Crystal based Quantum Memory for Light}.
\newblock PhD thesis, University of Aarhus Institute for Physics and Astronomy,
  2005.

\bibitem{Drewsen98}
M.~Drewsen, C.~Brodersen, L.~Hornek{\ae}r, and J.S. Hangst.
\newblock Large ion crystals in a linear paul trap.
\newblock {\em Physical Review Letters}, 81:2878--2881, 1998.

\bibitem{Mortensen06}
A.~Mortensen, E.~Nielsen, T.~Matthey, and M.~Drewsen.
\newblock Observation of three-dimensional long-range order in small ion
  coulomb crystals in an rf trap.
\newblock {\em Physical Review Letters}, 96:103001, 2006.

\bibitem{Madsen00}
D.N. Madsen, S.~Balslev, M.~Drewsen, N.~Kj{\ae}rgaard, Z.~Videsen, and J.W.
  Thomsen.
\newblock Measurements on photo-ionization of {$3s3p \; {}^1P_1$} magnesium
  atoms.
\newblock {\em Journal of Physics B}, 33:4981--4988, 2000.

\bibitem{Herskind07}
P.~Herskind, J.~Lindballe, C.~Clausen, J.L. S{\o}rensen, and M.~Drewsen.
\newblock Second-harmonic generation of light at 544 and 272 nm from an
  ytterbium-doped distributed-feedback fiber laser.
\newblock {\em Optics Letters}, 32:268--270, 2007.

\bibitem{Bolovinos96}
A.~Bolovinos et~al.
\newblock {$4pnp \; J = 0^e-2^e$} autoionizing series of calcium: experimental
  and theoretical analysis.
\newblock {\em Zeitschrift f{\"u}r Physik D}, 38:265--277, 1996.

\bibitem{Raab87}
E.L. Raab, M.~Prentiss, A.~Cable, S.~Chu, and D.E. Pritchard.
\newblock Trapping of neutral sodium atoms with radiation pressure.
\newblock {\em Physical Review Letters}, 59:2631--2634, 1987.

\bibitem{Reichel02}
J.~Reichel.
\newblock Microchip traps and {B}ose-{E}instein condensation.
\newblock {\em Applied Physics B}, 75:469--487, 2002.

\bibitem{Bradley00}
C.C. Bradley, J.J. McClelland, W.R. Anderson, and R.J. Celotta.
\newblock Magneto-optical trapping of chromium atoms.
\newblock {\em Physical Review A}, 61:053407, 2000.

\bibitem{Griesmaier05}
A.~Griesmaier, J.~Werner, S.~Hensler, J.~Stuhler, and T.~Pfau.
\newblock Bose-{E}instein condensation of chromium.
\newblock {\em Physical Review Letters}, 94:160401, 2005.

\end{thebibliography}
%
%
\end{document}